\documentclass[jkps,preprint,fleqn,showkeys]{revtex4}
\usepackage{graphicx}
\usepackage{amssymb}
\usepackage{amsmath}
\usepackage{bm}
\begin{document}
\setcounter{page}{0}
\title[]{Lattice and Electronic properties of VO$_2$ with the SCAN(+$U$) approach}
\author{Sooran \surname{Kim}}
\email{sooran@knu.ac.kr}
%\thanks{Fax: +82-2-554-1643}
\affiliation{Department of Physics Education, Kyungpook National University, Daegu 41566, Korea}

%\date[]{Received 6 January 2021}

\begin{abstract}
Appropriate consideration of the electron correlation is essential to reproduce the intriguing metal-insulator transition accompanying the Peierls-type structural transition in VO$_2$. In the density functional theory-based approach, this depends on the choice of the exchange-correlation functional. Here, using a newly developed strongly constrained and appropriately norm (SCAN) functional, we investigate the lattice and electronic properties of the metallic rutile phase of VO$_2$ ($R$-VO$_2$) from the first-principles calculations. We also explored the role of the Coulomb correlation $U$. By adding $U$, we found that the phonon instability properly describes the Peierls-type distortions. The orbital-decomposed density of states presents the orbital selective behavior with the SCAN+$U$, which is susceptible to the one-dimensional Peierls distortion. Our results suggest that even with the SCAN functional, the explicit inclusion of the Coulomb interaction is necessary to describe the structural transition of VO$_2$.   
\end{abstract}


\maketitle

\section{INTRODUCTION}

% VO2 structural transition explanation
Vanadium dioxide, VO$_2$ is one of the most intensively studied transition metal oxides due to its interesting metal-insulator transition accompanied by a structural transition at 340 K \cite{Mortin59,Goodenough60}.  
VO$_2$ is crystallized in the rutile structure ($R$-VO$_2$) with metallic behavior at high temperature \cite{McWhan74}. At a low temperature below 340 K, it exhibits  an insulating monoclinic phase ($M_1$-VO$_2$) \cite{Longo70}. Figure \ref{str} illustrates the two crystal structures of VO$_2$. The key distortion in the structural transition is related to V atoms having dimerization along the $c$-axis of $R$-VO$_2$ and zigzag distortion, which are typical Peierls distortion \cite{Eyert02,Pintchovski78}. 

% Previous works to explain transitions PRB+phonon
There has been an extensive discussion on the strong interplay between electron correlations and structural transition in VO$_2$ \cite{Biermann05,Lazarovits10,Kim13,Haverkort05}. To explain the phase transition and insulating behavior of $M_1$-VO$_2$, theoretical approaches beyond the density functional theory (DFT) level such as DFT+$U$ \cite{Liebsch05,Kim13} (the on-site Coulomb interaction $U$), $GW$ \cite{Continenza99,Gatti07,Sakuma08}, hybrid functional\cite{Eyert11,Budai14}, and the dynamical mean-field theory \cite{Biermann05,Tomczak08,Lazarovits10,Belozerov12} have been employed. Not only the electronic structures but also harmonic \cite{Kim13,Budai14,Mellan19} and anharmonic \cite{Budai14,Lee17} phonon properties were reported within the generalized-gradient approximation (GGA) and GGA+$U$ frameworks. Especially, Kim \textit{et al.} suggested that the Coulomb correlation $U$ plays an essential role in producing the Peierls-type structural transition \cite{Kim13}. 

% SCAN functional
 Previous theoretical works demonstrate that appropriate consideration of the electron correlation is essential for the sound description of the electronic and structural transition of VO$_2$.
 Recently, a new strongly constrained and appropriately normed (SCAN) functional was proposed, which is a non-empirical meta-GGA introducing the kinetic energy density and satisfying all known exact constraints \cite{Sun15,Sun16}. This new functional shows the improved performance on energetics and structural properties of binary oxides \cite{Hinuma17} and perovskite ferroelectrics \cite{Zhang17} as well as the band gaps and absolute voltages of cathode materials \cite{Chakraborty18} over GGA(+$U$) functional. Since SCAN does not require a tunable parameter $U$ to explain key properties of typical transition metal oxides, employing the SCAN functional increases the predictive power while keeping the first-principles character from the view of the \textit{ab initio} community \cite{Varignon19}. SCAN, however, also has an intrinsic self-interaction error as in the GGA functional \cite{Zhang18}, particularly from the local Coulomb interaction. The Coulomb interaction $U$ is often introduced in the SCAN to reduce the error \cite{Isaacs20,Gautam18,Olivia20}. Therefore, the SCAN functional needs to be carefully tested for each property and phase of different materials. 
 
 For VO$_2$, there have been previous studies of the energetics and electronic structures using the SCAN \cite{Ilkka17,Olivia20,Stahl20,Ganesh20,mondal20}. Notably, Mondal \textit{et al.} reported vibrational properties of VO$_2$ with SCAN functional \cite{mondal20}. Their phonon band of $R$-VO$_2$ with the SCAN is almost same as using GGA functional and does not exhibit an evidence for the Peierls-like structural distortion \cite{mondal20}. Therefore, considering an ongoing discussion on the SCAN, it would be worth exploring the lattice properties of the high-temperature phase, $R$-VO$_2$ with the SCAN(+$U$) approach.

In this paper, we investigate the lattice dynamics and electronic structure of $R$-VO$_2$ to demonstrate the performance of the newly proposed SCAN functional on VO$_2$. We also considered various $U$ values to examine the effect of Coulomb interaction with the SCAN. From the phonon dispersion curves with the SCAN and the SCAN+$U$, we find that phonon soft modes, which implies the structural instability, are significantly changed by adding the Coulomb interaction. Specifically, the phonon soft mode with the SCAN+$U$ shows the lattice displacements that are related to the Peierls-type structural transition whereas the soft mode with the SCAN does not. Furthermore, we investigate the orbital-decomposed density of states and present the orbital redistribution with the inclusion of $U$, resulting in the Peierls-like distortion. \\

\section{COMPUTATIONAL METHOD}
All density functional theory (DFT) calculations were performed by the Vienna \textit{ab initio} simulation package (VASP) implementing pseudo potential band method \cite{Kresse96,Kresse962}. We employed SCAN \cite{Sun15,Sun16} as an exchange-correlation functional and further included $+U$ to account for the correlated $d$ orbitals of V atom within the Dudarev method \cite{Dudarev98}. The various $U_{eff}=U-J$ (0.0, 1.0, 2.0, 2.1, 2.2, 2.3, 2.5, 3.1 eV) are chosen for detailed investigation of the $U$ effect with the SCAN functional. The $U_{eff}$ of 2.5 eV is used for Figures unless otherwise specified. The energy cut for the plane waves and the \textbf{k}-point sampling are 520 eV and $8\times8\times12$ in the Monkhorst-Pack grid, respectively. The structures are fully relaxed including lattice parameters and atomic positions from the initial experimental $R$-VO$_2$ structure \cite{McWhan74}. 
 
 Phonopy was employed for phonon study where the dynamic matrix and the force constants are obtained with the finite displacements method based on the Hellmann-Feynman theorem \cite{Togo08}. The $2\times2\times2$ supercell of the $R$-VO$_2$ and the $4\times4\times6$ {\textbf k}-point sampling were used for the phonon calculations. 

\section{RESULTS AND DISCUSSION}

Figure \ref{phonon} shows the phonon dispersion curves of $R$-VO$_2$ with the SCAN and SCAN+$U$ approaches. Phonon bands for both cases present phonon soft modes with imaginary frequencies, indicating the structural instability. These instabilities are in agreement with the experimentally unstable $R$-VO$_2$ phase at the low temperature. The phonon soft modes are mostly related to the V atoms as shown in the phonon density of states (DOS). The detailed displacements by the phonon soft modes will be discussed later. The phonon bands, however, with the SCAN and SCAN+$U$ clearly show the discrepancy at the $q$-points where the instability occurs. The phonon soft modes are obtained at $\Gamma$, $M$, and $X$ without $U$ while the soft modes exist in the whole Brillouin zone except for the $\Gamma$ point with the SCAN+$U$. Such behaviors are similar to the previous reports with GGA and GGA+$U$ approaches\cite{Kim13}. The phonon dispersion curves with the GGA and GGA+$U$ exhibit the phonon soft modes at ($\Gamma$, $M$, $X$) and ($A$, $Z$, $R$), respectively \cite{Kim13}.

Since the phonon softening at the $R$ point was reported to explain the structural transition from $R$-VO$_2$ to $M_1$-VO$_2$ \cite{Brews70,Hearn72,Terauchi78,Gervais85,Kim13}, our phonon results suggest that, even with the SCAN functional, the inclusion of the Hubbard parameter $U$ is necessary to describe the structural transition of VO$_2$. Namely, SCAN without adding explicit $U$ could not fully include the electron correlation effect in VO$_2$. These results raise the question of when SCAN requires \textit{additional} correction such as  $U$ and van der Waals interaction, in the ongoing discussion about SCAN functional \cite{Olivia20,Gautam18,Isaacs20,Peng16,Hermann18,kim20}. 

To investigate the $U$ effect on the phonon bands in more detail, we have plotted the lowest phonon frequencies at $\Gamma$ and $R$ with varying the effective $U$, $U_{eff}$ as shown in Fig. \ref{GR}. As increasing $U_{eff}$, the phonon soft modes disappear at $\Gamma$ while the soft modes appear at the $R$ point. Specifically, the phonon soft mode at $R$ starts to occur with $U_{eff}\geq$ 2.1 eV. When $U_{eff}\geq$ 2.3 eV, the soft optical modes at $\Gamma$ disappear and the lowest phonon mode at $\Gamma$ is the acoustic mode with a frequency of 0 meV. The lowest phonon modes at $A$ and $Z$ exhibit a similar trend with the soft modes at $R$. These results suggest that at least $U_{eff}$ of $\sim$2.5 eV is required to reproduce the structural Peierls transition in VO$_2$ within SCAN+$U$ approach.  

The displacements by the lowest phonon soft modes at $\Gamma$ with the SCAN and $R$ with the SCAN+$U$ are illustrated in Fig. \ref{dis}(a) and (b), respectively. The distortions at both $\Gamma$ and $R$ points are related to V atoms as shown in the phonon DOS of Fig. \ref{phonon}. However, the detailed displacements are different from each other. The phonon mode at $\Gamma$ exhibits collinear displacements of atoms along the $c$-axis while the corresponding displacement at $R$ shows the simultaneous dimerization and zigzag distortion of V atoms. The ordering of the dimerization by the phonon soft mode at $R$ also agrees well with the experimental distortion. In detail, there are two degenerated phonon soft modes at $R$ with the SCAN+$U$ as shown in Fig. \ref{dis}(c) and (d). One can see the dimerization in half of V atoms and zigzag distortion in another half of V atoms in each mode. By the linear combination of two modes, the dimerization and zigzag distortion in V ions are obtained at the same time as in Fig. \ref{dis}(b). Again, it suggests that the SCAN+$U$ framework is required to describe the experimental structural distortion. The phonon soft mode with the lowest frequency at $A$ and $Z$ has the displacement of V-V dimerization but not zigzag distortion. The lattice displacements by the lowest phonon soft modes at $M$ and $X$ are related to neither dimerization nor zigzag distortion.

We further investigate the electronic DOS of $R$-VO$_2$ within the SCAN(+$U$) functional. Figure \ref{dos} shows the total and partial DOS of $R$-VO$_2$. Both DOS with the SCAN and SCAN+$U$ exhibit the metallic states in agreement with the experiments. The inclusion of $U$ does not notably change the hybridization between V and O atoms as in Fig.\ref{dos} (a) and (b). However, the orbital-decomposed DOS in Fig. \ref{dos}(c) and (d) show a clear difference between SCAN and SCAN+$U$. The occupation of $d_{x^2-y^2}$ whose lobe is along with the V-V dimer significantly increases with the inclusion of $U$. This pronounced one-dimensional orbital distribution can result in the Peierls-type distortion. Furthermore, the orbital selective character with adding $U$ is consistent with the phonon soft modes at $R$, $A$, and $Z$ points and was also reported in the previous GGA+$U$ calculations\cite{Kim13}.

%=========+=========+=========+=========+=========+=========+=========+=========

\section{CONCLUSIONS}

In conclusion, we investigate the lattice and electronic properties of the metallic rutile phase of VO$_2$ and demonstrate that the direct inclusion of the Coulomb interaction $U$ in the SCAN functional is essential to reproduce the experimental structural transition in VO$_2$. 
The phonon bands with the SCAN+$U$ exhibit the phonon soft mode at $R$ whose lattice displacement is the dimerization and zigzag distortion of V atoms in agreement with the experiments. On the other hand, the phonon soft modes with the SCAN does not produce such displacements.  
Furthermore, the SCAN+$U$ calculation provides the orbital-selective distribution of V $d$ orbitals, which is suitable for the Peierls-type distortion and consistent with the phonon results.
We hope that this work can stimulate further study on the performance of the SCAN functional for transition metal oxides.

%=========+=========+=========+=========+=========+=========+=========+=========

\begin{acknowledgments}

We thank Bongjae Kim and Kyoo Kim for helpful discussion. This research was supported by Kyungpook National University Research Fund, 2018. 
\end{acknowledgments}

\begin{figure}[t!]
\includegraphics[width=10.0cm]{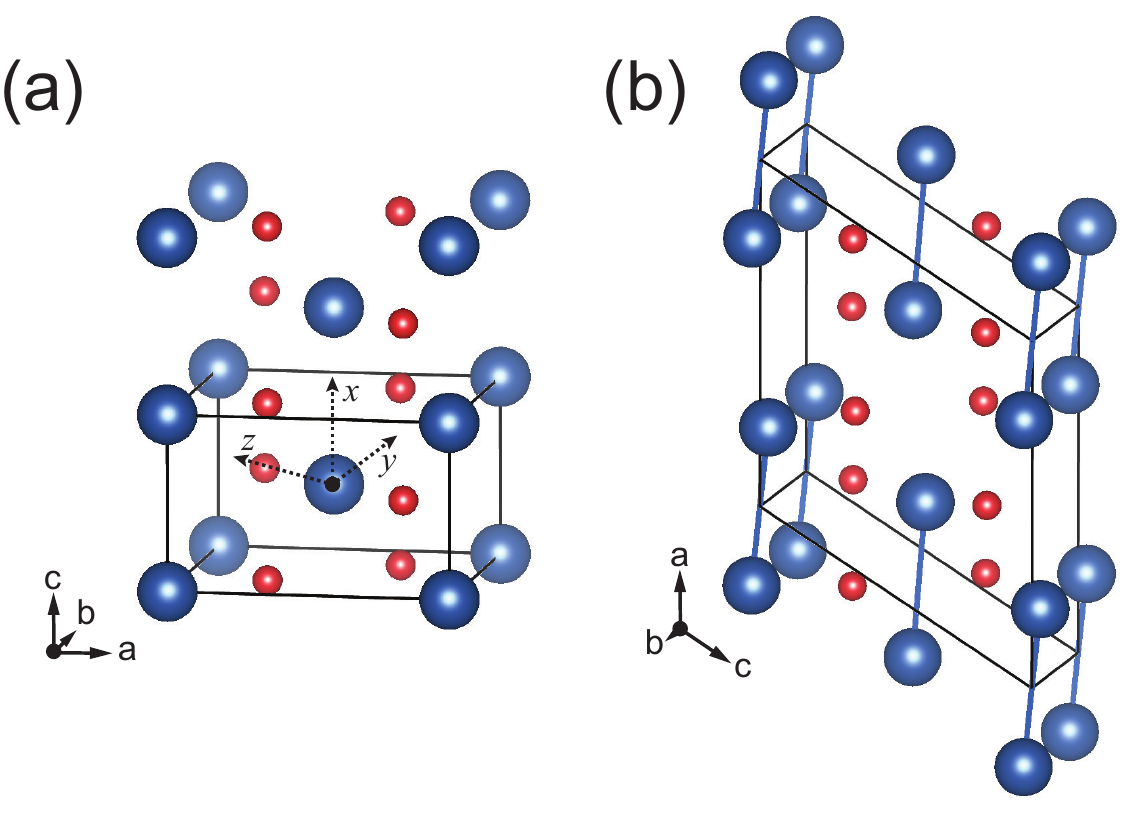}
\caption{} \label{str}
\end{figure}

\newpage
Figure \ref{str} Crystal structures of VO$_2$. (a) The rutile phase in the \textit{P}42/\textit{mnm} space group ($R$-VO$_2$) \cite{McWhan74}. The dotted arrow indicates the local axis following Ref. \cite{Eyert02} (b) The monoclinic phase in the \textit{P}21/\textit{c} space group ($M_1$-VO$_2$) \cite{Longo70}. The V-V dimerization and zigzag distortion in the monoclinic phase are indicated by the bond between V atoms. The blue and red balls represent V and O atoms, respectively.

\begin{figure}[t!]
\includegraphics[width=10.0cm]{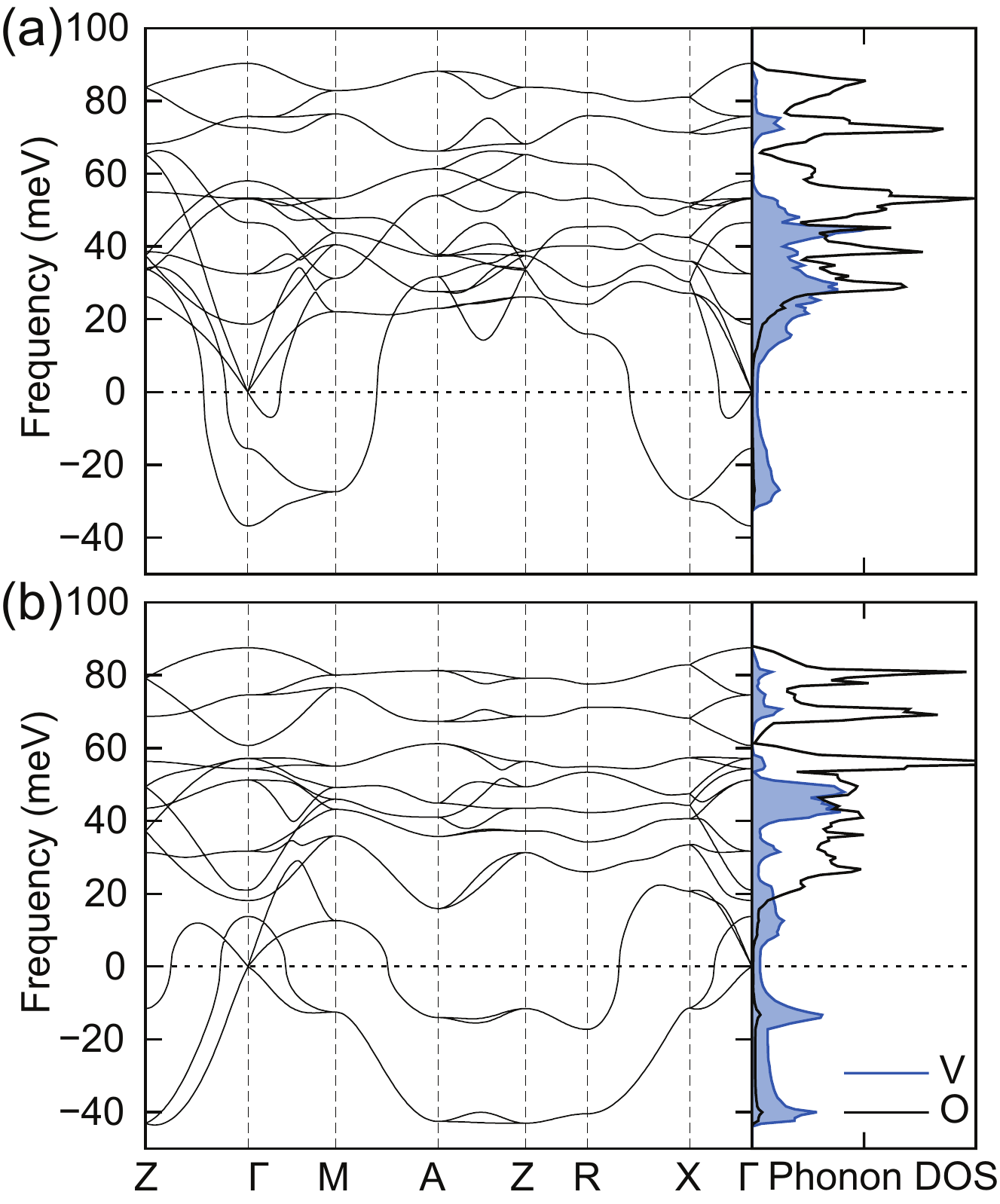}
\caption{} \label{phonon}
\end{figure}

\vspace{1 cm}
Figure \ref{phonon} Phonon dispersion curves and phonon density of states of $R$-VO$_2$ with (a) the SCAN and (b) the SCAN+$U$. The imaginary phonon frequencies in both figures imply the structural instability.

\begin{figure}[t!]
\includegraphics[width=10.0cm]{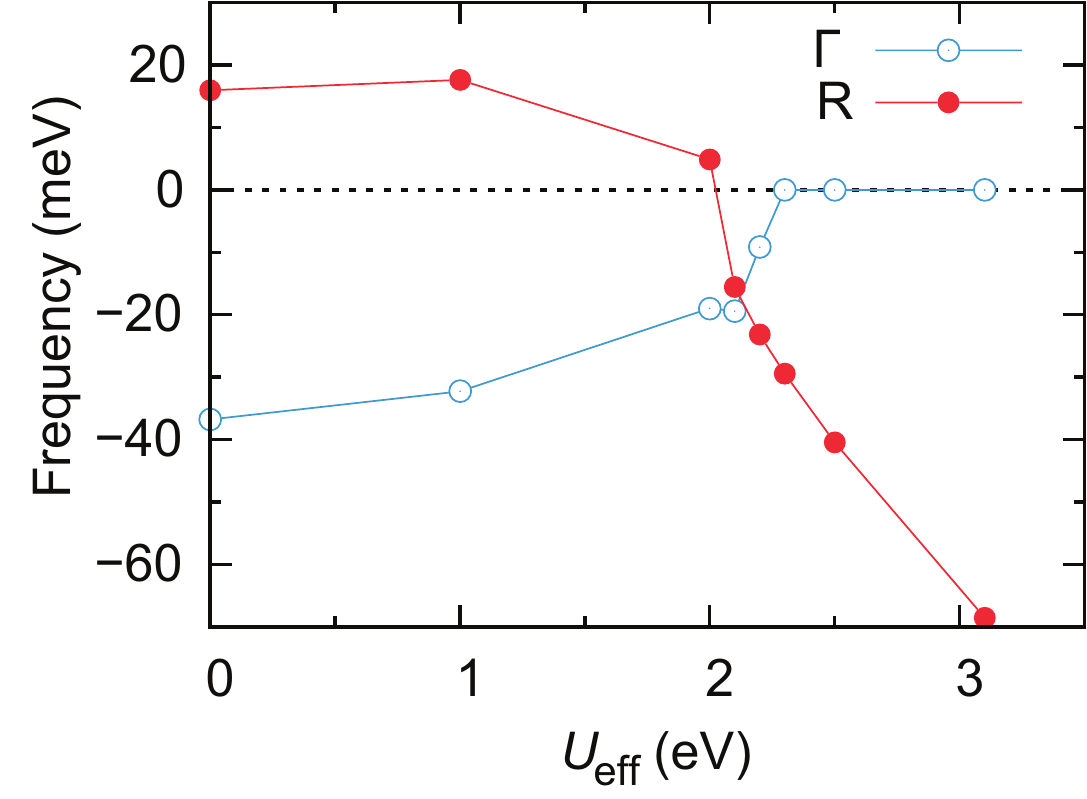}
\caption{} \label{GR}
\end{figure}

\vspace{1 cm}
Figure \ref{GR} The lowest phonon frequencies at $q=\Gamma$ and $R$ as a function of $U_{eff}$. 

\begin{figure}[t!]
\includegraphics[width=13.0cm]{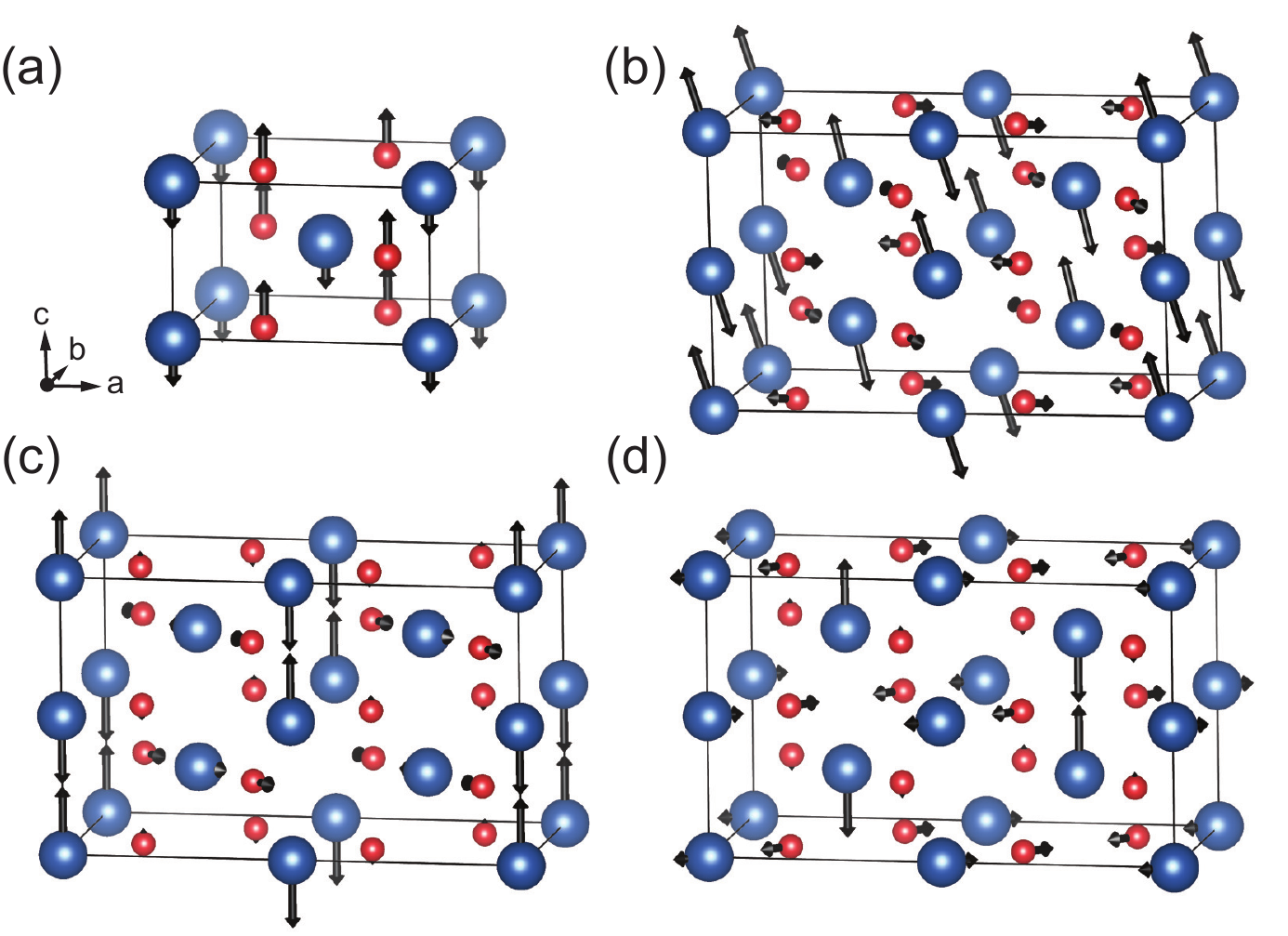}
\caption{} \label{dis}
\end{figure}

\vspace{1 cm}
Figure \ref{dis} (a) Lattice displacements by the phonon soft mode with the lowest frequency at $\Gamma$ with the SCAN. (b) A linearly superposed displacement by the lowest two degenerated phonon soft modes at $R$ as shown in (c) and (d) with the SCAN+$U$.

\begin{figure}[t!]
\includegraphics[width=13.0cm]{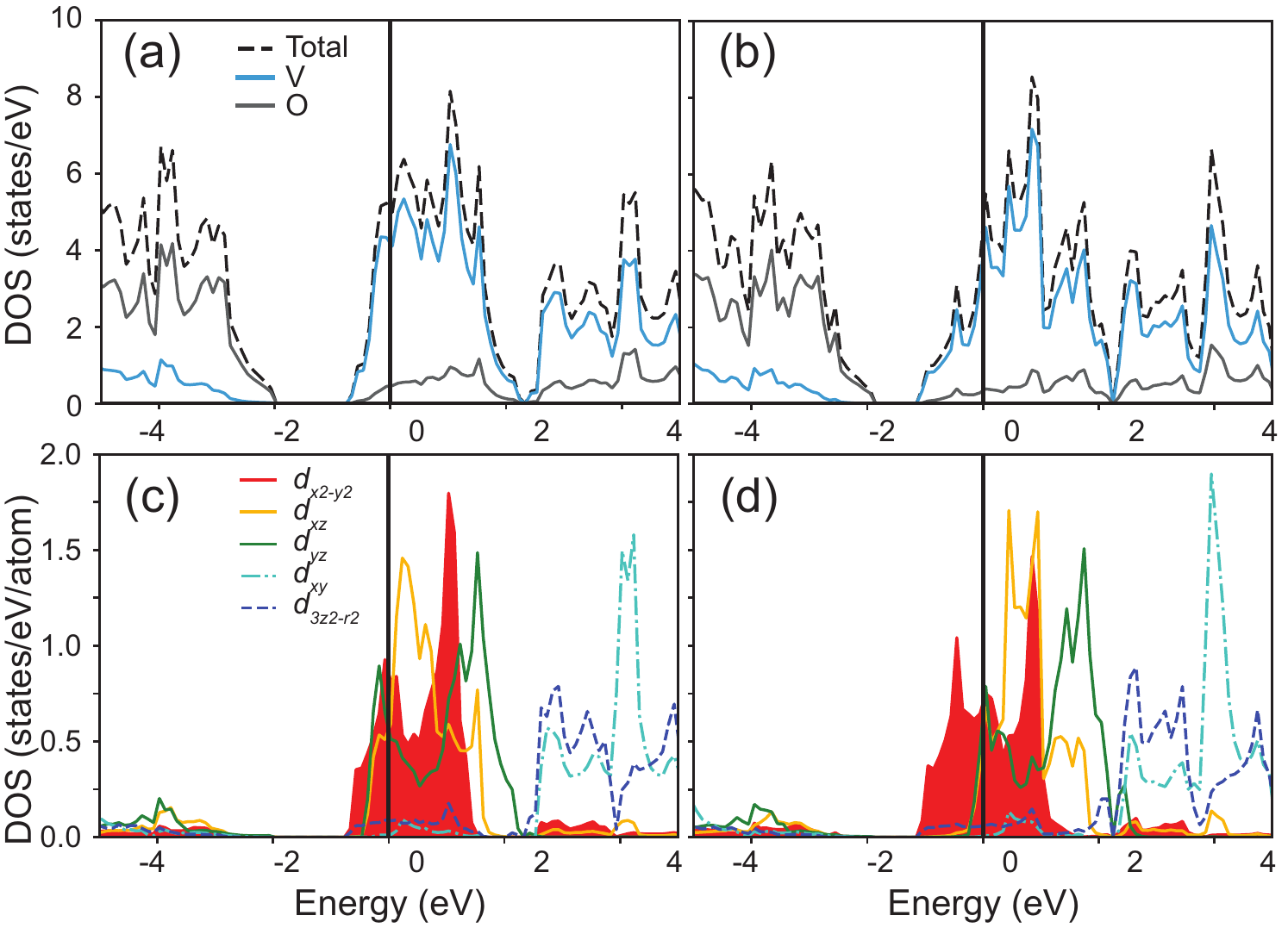}
\caption{} \label{dos}
\end{figure}

\vspace{1 cm}
Figure \ref{dos} The total and atomic decomposed DOS of $R$-VO$_2$ with (a) the SCAN and (b) the SCAN+$U$. The $d$ orbital-decomposed DOS of $R$-VO$_2$ with (c) the SCAN and (d) the SCAN+$U$.

\newpage

\newpage

\newpage

\newpage

\newpage

\end{document}